\documentclass[aps,prc,twocolumn,showpacs,10pt,dvipsnames,superscriptaddress]{revtex4-1}

\usepackage[utf8]{inputenc}
\usepackage{amsfonts}
\usepackage{amsmath}
\usepackage{amssymb}
\usepackage[pdftex]{graphicx}
\usepackage{xcolor}
\usepackage{tikz}
\usepackage{cancel}
\usetikzlibrary{arrows,patterns}
\usepackage{soul}

\begin{document}

\newcommand{\tj}[6]{ \begin{pmatrix}
  #1 & #2 & #3 \\
  #4 & #5 & #6
 \end{pmatrix}}


\title{$^9$Be+$^{120}$Sn scattering at near-barrier energies within a four body model}




\author{A. Arazi}
\email{arazi@tandar.cnea.gov.ar}

\affiliation{Laboratorio TANDAR, Comisi\'on Nacional de Energ\'{\i}a At\'omica, Av.
Gral. Paz 1499, BKNA1650 San Mart\'{\i}n, Argentina}
\affiliation{Consejo Nacional de Investigaciones Cient\'{\i}ficas y T\'ecnicas, Av. Rivadavia 1917, C1033AAJ Buenos Aires, Argentina}

\author{J. Casal}
\email{jcasal@us.es}
\affiliation{Departamento de F\'{\i}sica At\'omica, Molecular y Nuclear,
  Facultad de F\'{\i}sica, Universidad de Sevilla, Apartado 1065, E-41080
  Sevilla, Spain} 
\affiliation{European Centre for Theoretical Studies in Nuclear Physics and Related
Areas (ECT$^*$) and Fondazione Bruno Kessler, Villa Tambosi, Strada delle Tabarelle
286, I-38123 Villazzano (TN), Italy}

\author{M. Rodr\'{\i}guez-Gallardo}
\author{J. M. Arias}
\affiliation{Departamento de F\'{\i}sica At\'omica, Molecular y Nuclear,
  Facultad de F\'{\i}sica, Universidad de Sevilla, Apartado 1065, E-41080
  Sevilla, Spain} 
  
  \author{R. Lichtenth\"aler Filho}
\affiliation{Departamento de F\'{\i}sica Nuclear, Instituto de F\'{\i}sica da 
Universidade de S\~ao Paulo, 05508-090 S\~ao Paulo, SP, Brazil}
  
\author{D. Abriola}
 \author{O. A. Capurro}
  \affiliation{Laboratorio TANDAR, Comisi\'on Nacional de Energ\'{\i}a At\'omica, Av. Gral. Paz 1499, BKNA1650 San Mart\'{\i}n, Argentina}
 
 \author{M. A. Cardona}
 \author{P. F. F. Carnelli}
 
  \affiliation{Laboratorio TANDAR, Comisi\'on Nacional de Energ\'{\i}a At\'omica, Av. Gral. Paz 1499, BKNA1650 San Mart\'{\i}n, Argentina}
 \affiliation{Consejo Nacional de Investigaciones Cient\'{\i}ficas y T\'ecnicas, Av. Rivadavia 1917, C1033AAJ Buenos Aires, Argentina}
 
 \author{E. de Barbar\'a}
   \affiliation{Laboratorio TANDAR, Comisi\'on Nacional de Energ\'{\i}a At\'omica, Av. Gral. Paz 1499, BKNA1650 San Mart\'{\i}n, Argentina}

\author{J. Fern\'andez Niello}
 \author{J.~M.~Figueira}

  \affiliation{Laboratorio TANDAR, Comisi\'on Nacional de Energ\'{\i}a At\'omica, Av. Gral. Paz 1499, BKNA1650 San Mart\'{\i}n, Argentina}
 \affiliation{Consejo Nacional de Investigaciones Cient\'{\i}ficas y T\'ecnicas, Av. Rivadavia 1917, C1033AAJ Buenos Aires, Argentina}
 \author{L. Fimiani}
 
    \affiliation{Laboratorio TANDAR, Comisi\'on Nacional de Energ\'{\i}a At\'omica, Av. Gral. Paz 1499, BKNA1650 San Mart\'{\i}n, Argentina}
 
 \author{D. Hojman}
 
  \affiliation{Laboratorio TANDAR, Comisi\'on Nacional de Energ\'{\i}a At\'omica, Av. Gral. Paz 1499, BKNA1650 San Mart\'{\i}n, Argentina}
 \affiliation{Consejo Nacional de Investigaciones Cient\'{\i}ficas y T\'ecnicas, Av. Rivadavia 1917, C1033AAJ Buenos Aires, Argentina}

 \author{G. V. Mart\'{\i}}
 \affiliation{Laboratorio TANDAR, Comisi\'on Nacional de Energ\'{\i}a At\'omica, Av. Gral. Paz 1499, BKNA1650 San Mart\'{\i}n, Argentina}
 
\author{D. Mart\'{\i}nez Heimman}
 \author{A. J. Pacheco}
  
  \affiliation{Laboratorio TANDAR, Comisi\'on Nacional de Energ\'{\i}a At\'omica, Av. Gral. Paz 1499, BKNA1650 San Mart\'{\i}n, Argentina}
 \affiliation{Consejo Nacional de Investigaciones Cient\'{\i}ficas y T\'ecnicas, Av. Rivadavia 1917, C1033AAJ Buenos Aires, Argentina}

\date{\today}

\begin{abstract}
 Cross sections for elastic and inelastic scattering of the weakly-bound  $^9$Be nucleus on a $^{120}$Sn target have been
 measured at seven bombarding energies around and above the Coulomb barrier.
The elastic angular distributions are analyzed with a four-body
continuum-discretized coupled-channels (CDCC) calculation, which considers $^9$Be as
a three-body projectile ($\alpha$ + $\alpha$ + n). An optical model analysis using
the S\~ao Paulo potential is also shown for comparison. The CDCC analysis shows that
the coupling to the continuum part of the spectrum is important for the agreement
with experimental data even at energies around the Coulomb barrier,  suggesting that breakup is an important process at low energies. At the highest incident
energies, two inelastic peaks are observed at 1.19(5) and 2.41(5) MeV.
Coupled-channels (CC) calculations using a rotational model confirm that the first
inelastic peak corresponds to the excitation of the 2$_1^+$ state in $^{120}$Sn, 
while the second one likely corresponds  to the excitation of the 3$_1^-$
state.

\end{abstract}



\maketitle


\section{Introduction}

Weakly-bound and exotic nuclei have been intensively studied due to 
their role in astrophysics \cite{tho09} and as a test for theoretical models capable 
of describing their singular structure and complex reaction mechanisms
\cite{cluster}. This interest has been boosted by the availability of radioactive
ion beam facilities which allowed experimental studies of reactions involving these nuclei~\cite{blu13,lepine14,lichten16}. 
One of their main features is the breakup process, which is supposed to be triggered by the coulomb (nuclear) interaction when scattering on a 
heavy (light) target.  The breakup process may affect other reaction channels as fusion, and the assessment of this effect has been the subject of several theoretical and experimental efforts \cite{hin02,pak03,pak03b, pak04,fig06,sou07,gar07,kol07,man07,bec07,bis08,gom08,kee08,muk09,has09,kuc09,mon09,lub09,gom09, can09,gar09,sou09,zer12,cubero12,juanpi13,morcelle14}.
Nevertheless, this effect is still not totally clear and contradictory results
coexist.

 Experimentally, the breakup process can be studied by the detection in 
coincidence of all the breakup fragments \cite{sho81, hes91, 
gui00,kol01, sig03, shr06,pak06, pak07, mar14, san09,car17}. In many
cases, it requires neutron detection which can be rather involved on the 
experimental point of view. In addition,  the breakup of light 
projectiles usually produces fragments with masses and charges
similar to other light particles coming from different decay processes 
such as fusion, or even direct reaction channels such as transfers.
For these reasons it is not easy to unambiguously identify the breakup 
process.

These studies are extremely difficult and time demanding to be performed
with radioactive projectiles since they are produced as secondary beams with
intensities several orders of magnitude lower than stable projectiles. Hence,
stable weakly-bound nuclei, such as $^6$Li, $^7$Li, and $^9$Be, which are produced
as primary beams with regular intensities, offer an excellent opportunity to perform
systematic studies of angular distributions of their reaction products. 

On the other hand, a big theoretical effort has 
been made over the last decades to develop coupled channels calculations 
schemes to take into account the effect of the breakup
process on the elastic scattering angular distributions. Three-body and 
four-body Continuum discretized coupled channels calculations have been 
applied to a number of cases with great success \cite{net10,pir11,morcelle14}.

The nucleus of $^9$Be presents a Borromean structure comprising two $\alpha$
particles and one neutron. Although stable, $^9$Be  has a small binding energy  of
1.5736 MeV below the  $\alpha+\alpha+n$ threshold~\cite{Tilley04}. Therefore, when colliding with a target nucleus, breakup effects  are expected to be relevant. 
 Experimental efforts have been made to better determine the $^9$Be structure, such as, works in Refs.~\cite{fult04,ashw05,pap07,brown07}.
 Reactions induced by $^9$Be have been already studied on $^{208}$Pb at the
Australian National University~\cite{Wooll04} and at the China Institute of
Atomic Energy~\cite{Yu10}, on $^{27}$Al at the University of S\~ao Paulo and on $^{27}$Al and $^{144}$Sm at the TANDAR Laboratory~\cite{Gomes04,gom09}. 
Regarding the target, the spherical (proton magic) $^{120}$Sn nucleus has been investigated with weakly bound projectiles, both stable ($^{6,7}$Li \cite{sou10}) and radioactive ($^6$He \cite{net10} and $^8$Li \cite{net09}).   

The experimental data for elastic and breakup fragments, produced in reactions
involving these projectiles, can be compared with continuum-discretized
coupled-channels (CDCC) calculations \cite{Yahiro86,Austern87}, which include the
coupling to the continuum part of the spectrum or breakup channels
\cite{Matsumoto06, MRoGa08, JALay10, juanpi13}. 
The CDCC formalism, first developed for two-body projectiles (three-body CDCC), was
later extended to three-body projectiles (four-body
CDCC)~\cite{Matsumoto06,MRoGa08}. Very recently, the latter has been applied to 
$^9$Be-induced reactions~\cite{Descouvemont15,JCasal15}, taking into account its
Borromean structure.
 In Ref.~\cite{JCasal15}, the scattering of $^9$Be on $^{208}$Pb and $^{27}$Al at
energies around the Coulomb barrier was studied showing that Coulomb breakup 
is still important at this energy range. The relevance of the $^9$Be low-energy
resonances on the angular cross sections was also shown.

 In order to analyze the inelastic distributions due to the target excitation in the scattering of a weakly-boud projectile, it would be desirable to include such excitations consistently within the CDCC formalism.
Very recently, this extension has been  addressed for the case of the three-body CDCC (i.e., for a two-body projectile)~\cite{gomez17}. The feasibility of a similar approach for the four-body CDCC (three-body projectile) still needs to be studied. 
Nonetheless, Coupled-Channels (CC) calculations with collective form factors~\cite{tamura65} can be performed by including explicitly the most important states and a bare potential to reproduce the interaction between projectil and target in the absence of coupling to the internal degrees of freedom.

In this work we present new measurements for the scattering of $^9$Be on the
intermediate-mass target $^{120}$Sn carried out at the TANDAR laboratory. 
In section \ref{exp}, the experimental setup is addressed and the data is presented. In
section \ref{calc}, the measured elastic angular distributions are compared with an
optical model (OM) analysis using the S\~ao Paulo potential (SPP) and with the four-body CDCC
calculations. In section \ref{inel}, the experimental inelastic distributions are
briefly analyzed  with simple CC calculations. The summary and conclusions are given in section \ref{concl}.

\section{Experimental setup and results}
\label{exp}

The experimental elastic and inelastic scattering angular distributions were obtained at the 20~UD
tandem accelerator TANDAR at Buenos Aires. The $^9$Be beams were mostly extracted as
$^9$BeO$^-$ ions from the sputtering ion source, since their intensity (up to 1
$\mu$A at the ion source) is about 50 times higher than for atomic $^9$Be$^-$ ions.
For the lower energies ($E_{\rm lab}$= 26, 27, 28, 29.5, and 31~MeV), the 3+ charge state was selected, achieving a mean analyzed $^9$Be intensity of 15~pnA. Since the accelerator
was limited to a terminal voltage of 10~MV at the time of the experiment, the charge
state 4+ was tuned for $E_{\rm lab}=42$~MeV, yielding 5~pnA. To achieve $E_{\rm
lab}=50$~MeV, $^9$Be$^-$ ions were injected. In spite of their lower output at the
ion source, $^9$Be$^-$ ions have a much better transmission at the stripper, since
they do not suffer from the defocusing effect of the Coulomb explosion as the
$^9$BeO$^-$ molecular beam. Besides, $^9$Be$^-$ ions have a higher yield for the 
$q=4+$ charge state, achieving an analyzed intensity of 1~pnA.  

Targets of enriched ($>$99\%) $^{120}$Sn, 85~$\mu$g/cm$^2$ thick, evaporated onto
20~$\mu$g/cm$^2$ carbon foils, were used at the center of a 70-cm-diameter
scattering chamber. 
For some
energies ($E_{\rm lab}$ = 29.5, 42, and 50~MeV) a stack of two targets were used to
increase the counting rate. The energy loss in the target was calculated and the
energy in the center of it was assumed as the reaction energy.

An array of eight surface-barrier detectors (150~$\mu$m thick), with an angular
separation of 5$^\circ$ between adjacent detectors, was used to distinguish scattering 
products. A liquid nitrogen cooling system sets the detector temperature at $-$20$^\circ$C
to improve their energy resolution, which varied between 0.5\% and 1.0\% (FWHM). 
This allowed to separate two inelastic-excitation peaks with excitation energies of 
1.19(5)  and 2.41(5) MeV from the elastic-scattering peak.

\begin{figure}
  \centering
  \includegraphics[width=0.95\linewidth]{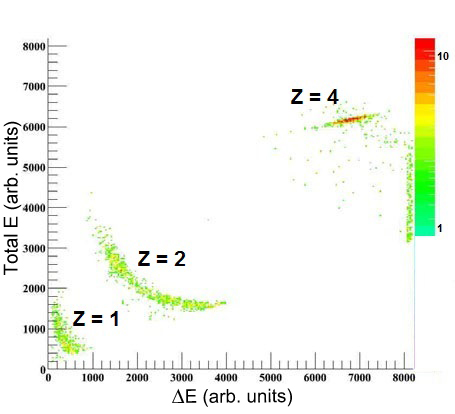}
  \caption{(Color online) Two-dimensional spectrum recorded at $\theta_{\rm lab} =
170$$^\circ$ for $E_{\rm lab} = 26$~MeV. The horizontal axis is the energy loss in the
first stage of the telescope  ($\Delta E$) while the vertical one is the total energy obtained as $E=\Delta E + E_{\rm res}$. The projection on this axis is
equivalent to one-dimensional spectra obtained by single detectors. The $Z$~=~1,
$Z$~=~2, and $Z$~=~4 groups can be clearly identified.}
\label{spectrum-2D}
\end{figure}

 The detectors were collimated by rectangular slits, defining an angular acceptance  smaller than 0.5$^\circ$ and solid angles varying between 0.07~msr (most forward detector) and 0.8~msr (most backward one). This assured comparable counting rates in all detectors.
Additionally, a silicon telescope detector $\Delta E$ (15 $\mu$m) - $E_{\rm res}$  (150 $\mu$m) was placed at 170$^\circ$. 
 The $E-\Delta E$ two-dimensional spectra (see Fig.~\ref{spectrum-2D}) allowed us  to evaluate the composition of the background (mainly alpha particles arising from the projectile breakup) at angles in which the energy and counting rate of the elastic scattering are the lowest. It can be seen that alpha particles have lower energies than the elastically and inelastically scattered $^9$Be and, therefore, they produced no interfering background, not even in the single detectors of our array, which  only measure  the total energy.  Simulations performed with the code SUPERKIN \cite{hei10}, assuming the same relative energy for the breakup fragments as observed at 170$^\circ$, allowed us to extend this result to forward angles. 
 
\begin{figure}
  \centering
  \includegraphics[width=0.95\linewidth]{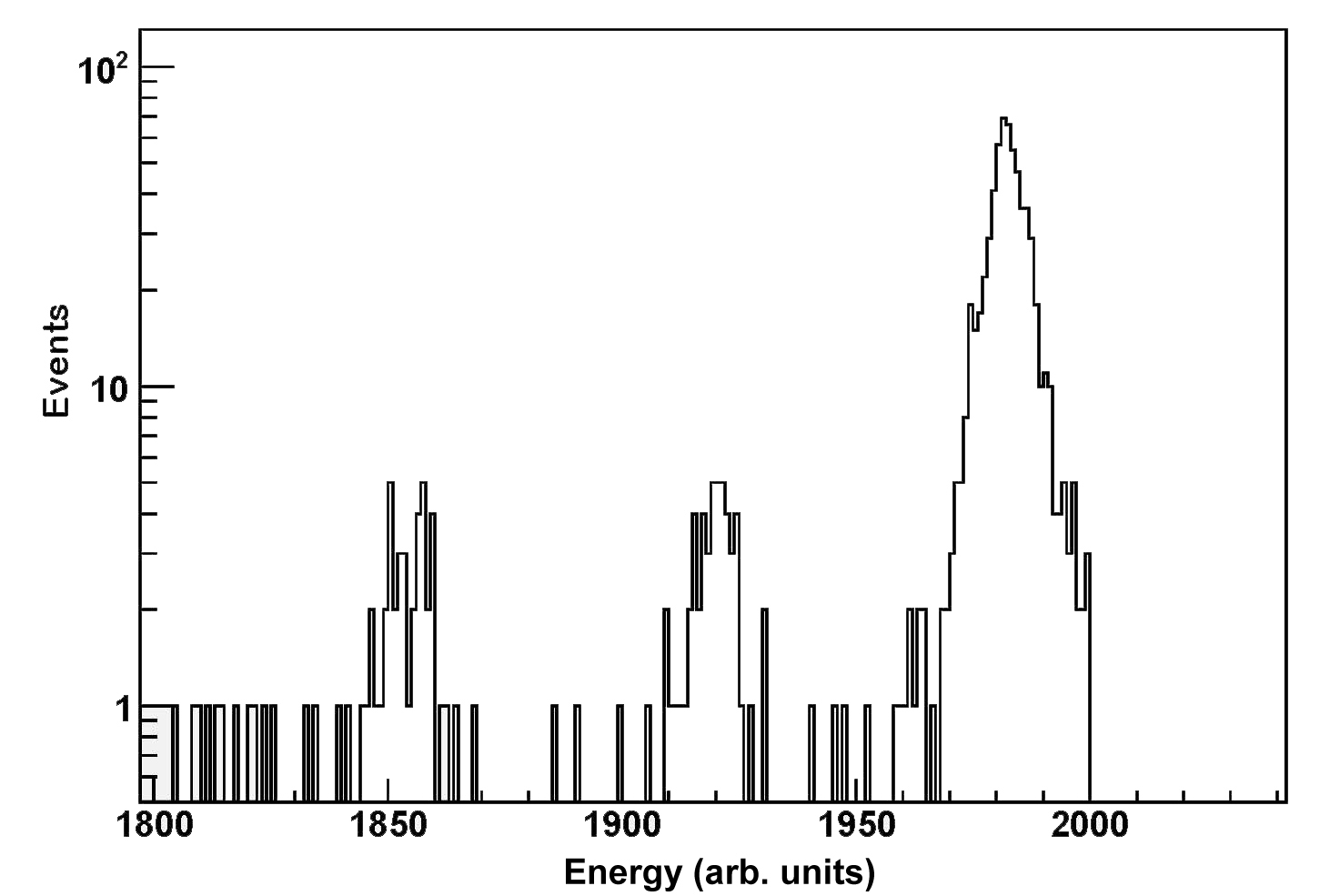}
  \caption{ Typical spectrum recorded at $\theta_{\rm lab} = 62.5$~deg for $E_{\rm
lab} = 42$~MeV. The large peak is due to the elastic scattering whereas the small 
ones correspond to inelastic scattering processes.} 
\label{spectrum-1D}
\end{figure}

  A typical one-dimensional spectrum ($\theta_{\rm lab} = 62.5$~deg,  $E_{\rm lab} =
42$~MeV)  is shown in Fig. \ref{spectrum-1D}. For the peak integration, an
asymmetric Gaussian curve was fitted to the histograms  with the lower (upper)
integration limit calculated as $x_{\rm low(upp)} = x_0-(+)3\sigma_{\rm low(upp)}$, where
$x_0$ is the centroid and $\sigma _{\rm low(upp)}$ is the lower (upper) standard
deviation.
 
The measured angular range extended from 22$^\circ$ to 170$^\circ$ (laboratory frame) except at
the highest energies where the angular range was progressively reduced. For $E_{\rm
lab} = 50$~MeV, the covered angular range was from 10$^\circ$ to 75$^\circ$. 

The normalization of cross sections was performed using a monitor detector  which remained at a fixed angle $\theta_{\rm mon}=25^\circ$, where the scattering is pure Rutherford. The differential cross sections for the $i-$th detector at angle position $\theta_i$, was then determined as
\begin{equation}
\label{x-section formula}
\frac{d\sigma}{d\Omega}(\theta_i)= 
\frac{d\sigma^{Ruth}}{d\Omega}(\theta_{\rm mon}) 
\frac{N_i}{N_{\rm mon}}
\frac{J_i}{J_{\rm mon}}
\frac{\Omega_{\rm mon}}{\Omega_i},
\end{equation} 
where $N_{i({\rm mon})}$, $J_{i({\rm mon})}$, and $\Omega_{i({\rm mon})}$ are the number of events in the peak, the Jacobian for the laboratory to center of mass transformation and the solid angle of the $i-$th detector(monitor), respectively. For determining the ratios
between the solid angles of each detector and that of the monitor, several angular distributions were measured with the detector array placed at different angles (both forward and backward) for two systems at sub-Coulomb energies, $^6$Li + $^{197}$Au at $E_{\rm lab}$ = 19 and 23~MeV and $^{16}$O$+^{197}$Au at $E_{\rm lab}  = 58$~MeV, for which the Rutherford angular distribution was assumed.  

An independent normalization was given by a Faraday cup at the end of the beam line,
far away from the target, which integrated the total charge delivered by the beam in
each run. 

The uncertainties in the cross section values were estimated as the root of the sum of squares of:  a) the statistical contribution from both detector and monitor counts, which was 
about 2\% on average (7\% maximal) for lower energies, except at the higher energies and backward angles, where it reached values of 20\% or 30\% due to the low counting, 
b) differences between the cross section values yielded by the normalization with the
Faraday cup and with the monitor (less than 3\% in most cases), and c) 2\% for other uncontrolled uncertainty sources as detector angular position (known better than 0.1$^\circ$), beam deviation, peak integration, etc. Hence, the total uncertainties typically ranged from 3\% to 8\%, with the aforementioned exceptions.
  
\begin{figure}[h]
\includegraphics[width=0.9\linewidth]{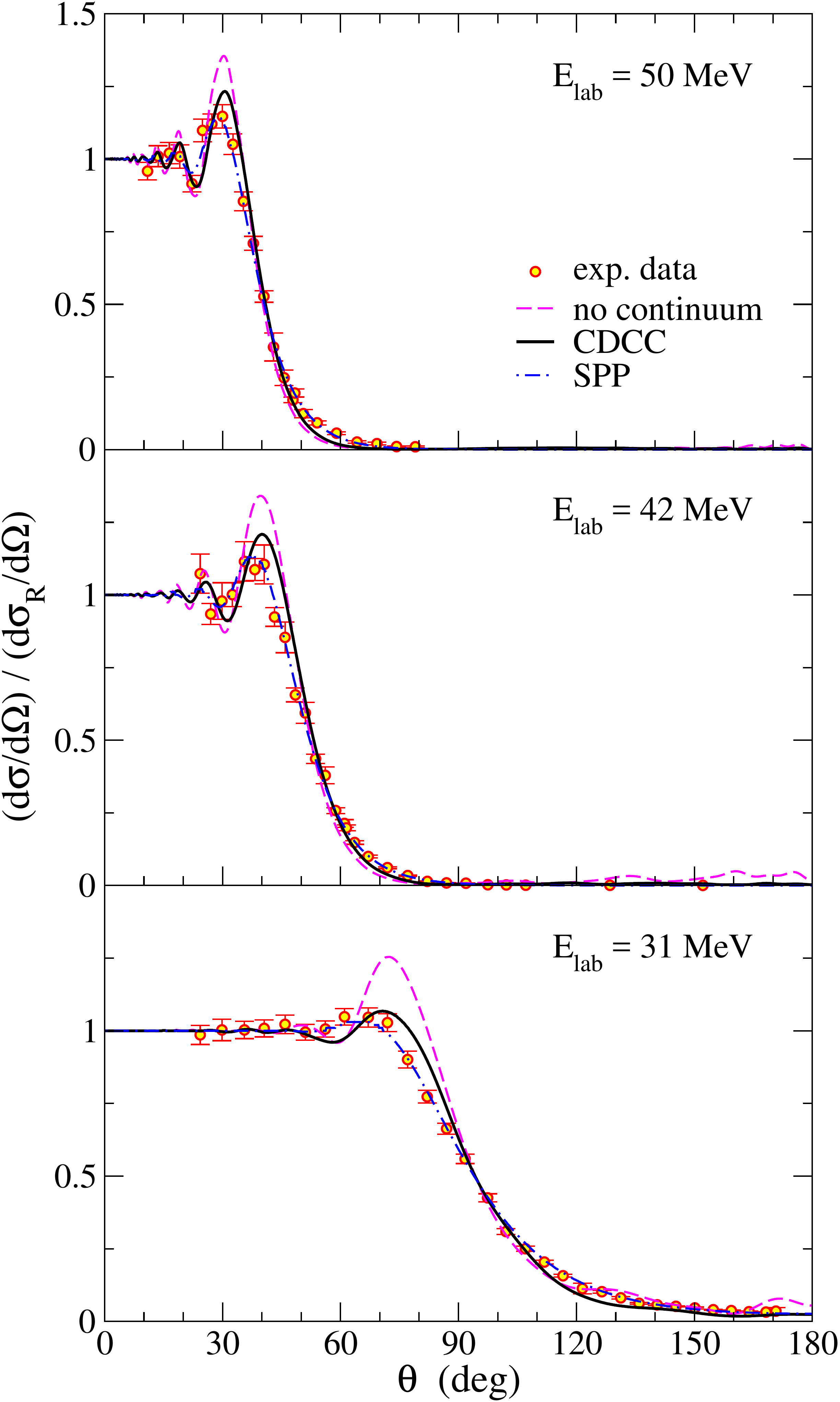}
\caption{(Color online) Angular distribution of the elastic cross section relative to Rutherford
for the reaction $^9$Be + $^{120}$Sn at $E_{\rm lab}=50, 42$ and 31 MeV. The present
experimental data are shown with circles. Dot-dashed lines correspond to optical model (OM)
calculations using the S\~ao Paulo potential (SPP). Dashed lines correspond
to four-body calculations including the ground state only, and solid lines show the
full four-body CDCC calculations.}
\label{fig:all1}
\end{figure}

\begin{figure}[h]
\raggedright
\includegraphics[width=0.9\linewidth]{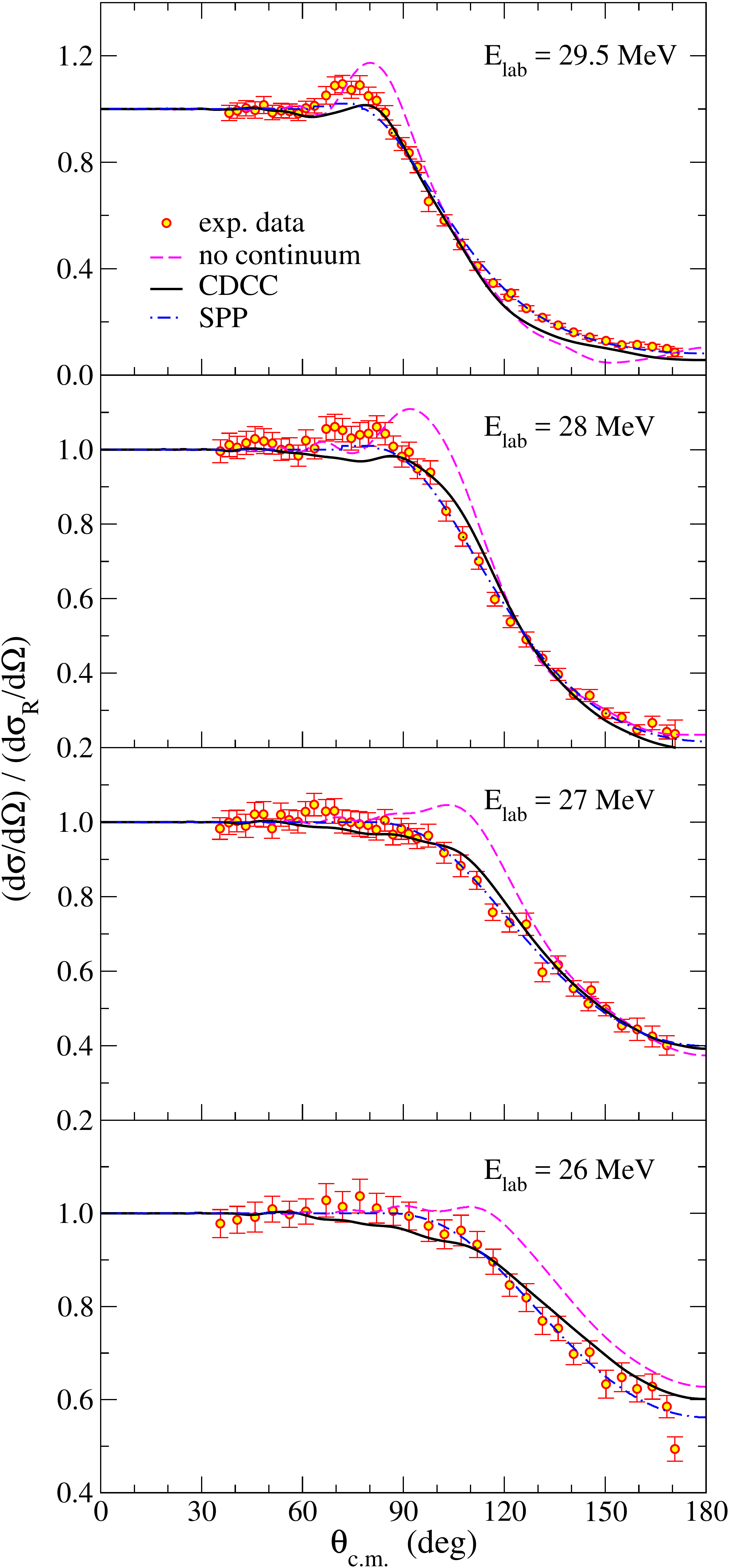}
\caption{(Color online) Angular distribution of the elastic cross section relative to Rutherford
for the reaction $^9$Be + $^{120}$Sn at $E_{\rm lab}= 29.5, 28, 27$ and $26$ MeV.
The present experimental data are shown with circles. Dot-dashed lines correspond to optical model 
(OM) calculations using the S\~ao Paulo potential (SPP). Dashed lines
correspond to four-body calculations including the ground state only, and solid
lines show the full four-body CDCC calculations.}
\label{fig:all2}
\end{figure}

\begin{figure}[h!]
 \includegraphics[width=1.15\linewidth]{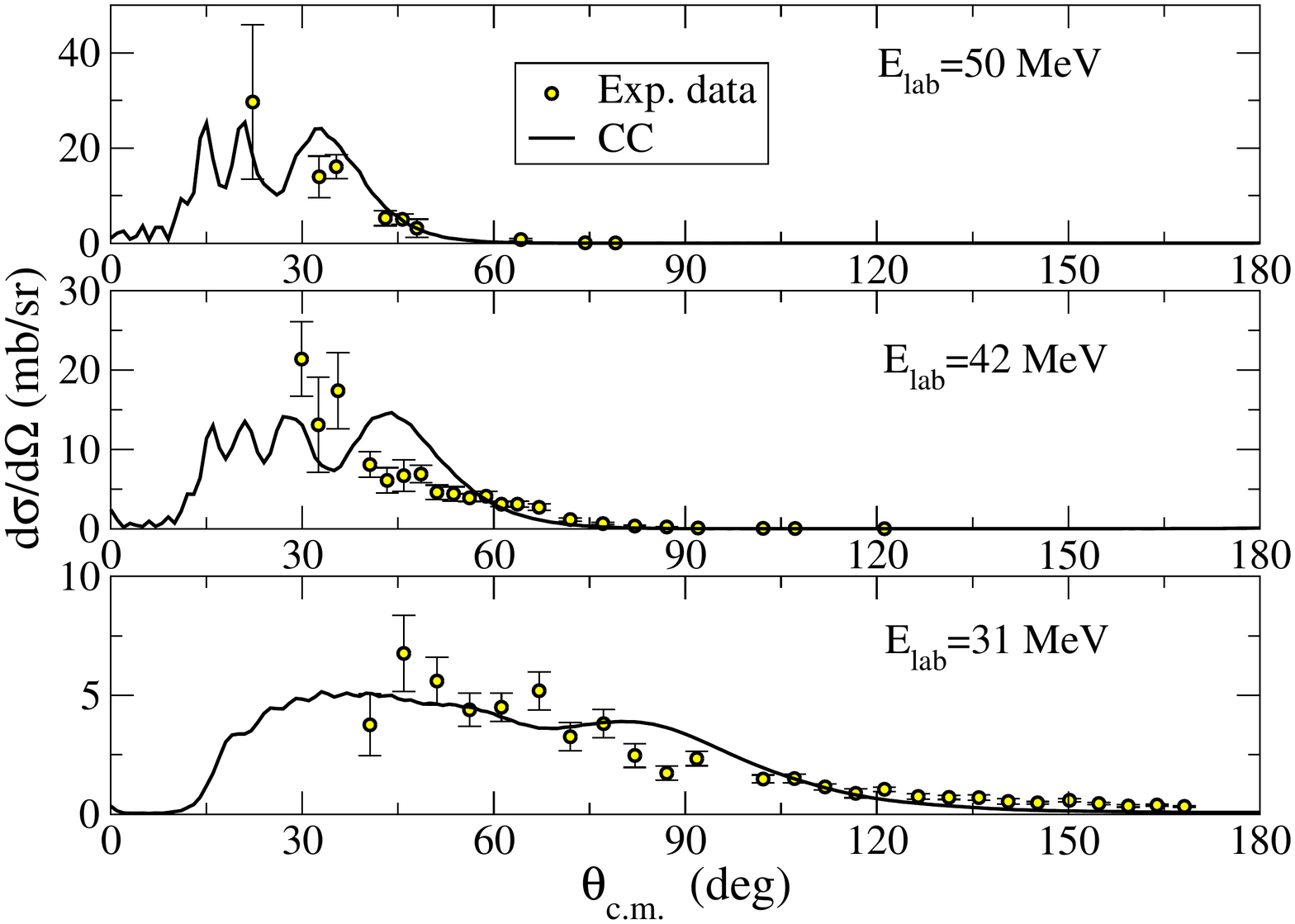}
 \caption{(Color online) Angular distribution of the inelastic cross section, considering an 
excitation of the $^{120}$Sn to its $2_1^+$ state, for the reaction  $^9$Be + $^{120}$Sn at 
$E_{\rm lab}=50$, 42, and 31~MeV.}
 \label{fig:inel1}
\end{figure}

\begin{figure}[h!]
 \includegraphics[width=1.15\linewidth]{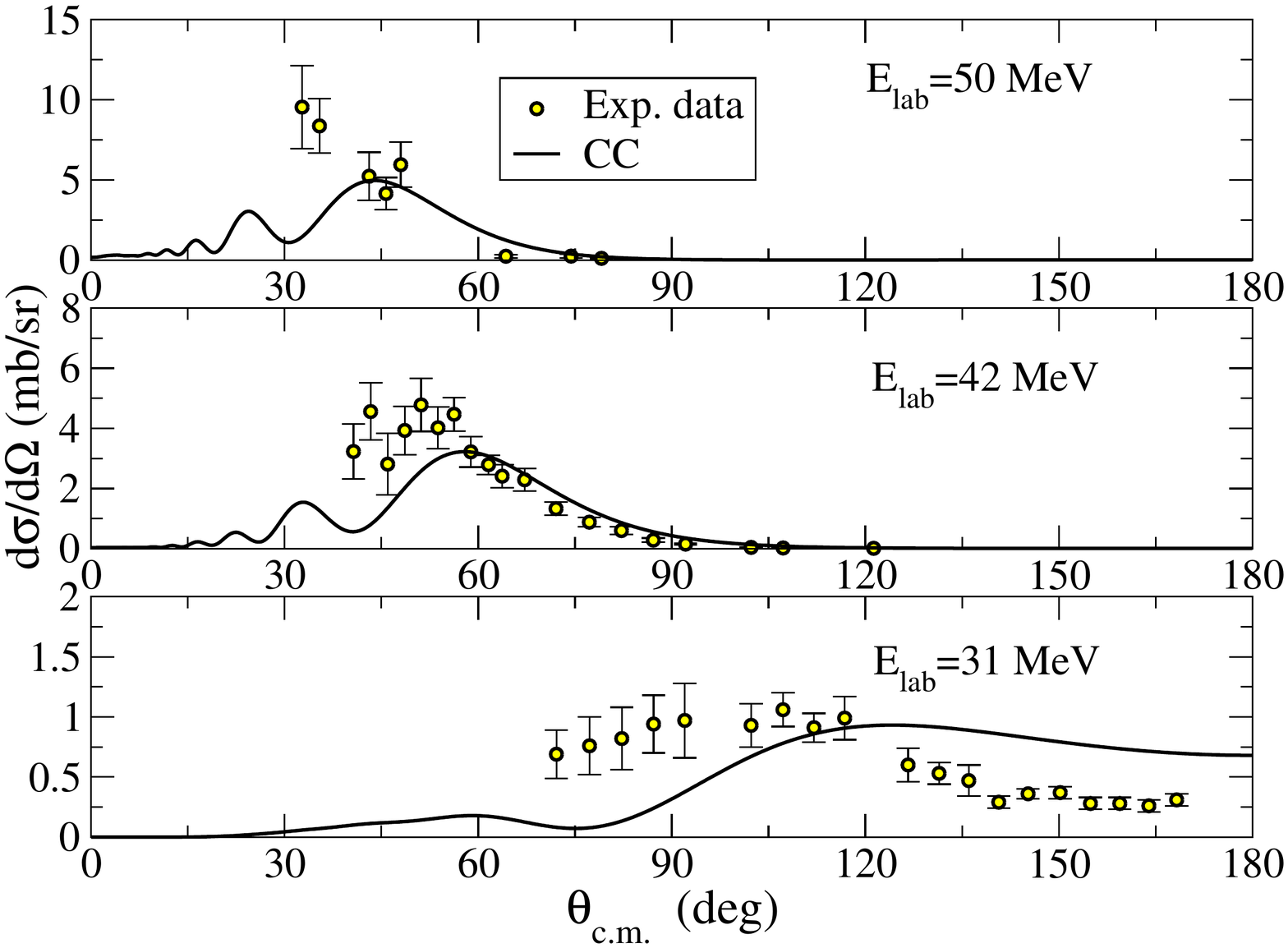}
 \caption{(Color online) Angular distribution of the inelastic cross section, considering an excitation of the  $^{120}$Sn to its $3_1^-$ state, for the reaction  $^9$Be + $^{120}$Sn at $E_{\rm lab}=50$, 42, 
and 31~MeV.}
 \label{fig:inel2}
\end{figure}

 The experimental angular distributions of the elastic-scattering cross sections
normalized to Rutherford cross section are shown in Fig.~\ref{fig:all1} for the
three highest energies measured ($E_{\rm lab}=50$, 42 and 31 MeV), and in
Fig.~\ref{fig:all2} for the other four energies ($E_{\rm lab}=29.5$, 28, 27 and 26
MeV). The inelastic cross-section distributions are shown in Fig.~\ref{fig:inel1}
and  Fig.~\ref{fig:inel2}, for the first and second peaks, respectively, at the
three highest energies (50, 42 and 31 MeV).

\section{Analysis of the elastic scattering}
\label{calc}

\subsection{Optical model analysis}
\label{OMA}

First, we performed optical model (OM) calculations using the S\~ao Paulo potential
(SPP)~\cite{chamon97,alv03}. In this model, the normalization factors of the real and the imaginary parts of the potential, $N_r$ and $N_i$ respectively, are obtained adjusting experimental elastic angular distributions at different bombarding energies. For the data of the present work, the best values obtained for these factors are presented in Table~\ref{OP-parameters}. The quality of the fit is
confirmed by  $\chi^2/\nu$ values which are close to unity. In Figs.~\ref{fig:all1} and \ref{fig:all2} the OM fit for each energy is shown with a dot-dashed line  and in Table~\ref{trxsec} the calculated total reaction cross section for each energy is displayed. 

 In the case of tightly bound nuclei, the imaginary factor $N_i$ drops at energies below the Coulomb barrier 
 (with a corresponding peak in the real factor $N_r$) and this effect is known as the threshold anomaly (TA). On the contrary, for some weakly bound nuclei $N_i$ has been found to increase below the Coulomb barrier ~\cite{pak03b,pak04,fig06, fig10,fim12}, which has been associated as the effect of the breakup channels still being open. This effect has been called breakup threshold anomaly (BTA)~\cite{hus06}. Intermediate cases in which neither behavior is clear have also been observed for $^7$Li \cite{fig06,fig10} and $^9$Be \cite{Gomes04} projectiles. For a global comparison using the same OM framework see \cite{zer12, abr15, abr17}.     

Concerning $^9$Be, there have been several OM calculations for different systems.  
The studies on the $^9$Be+$^{209}$Bi~\cite{sig00,Yu10,gom15} and $^9$Be+$^{208}$Pb~\cite{Yu10,gom15} systems suggest the presence of the BTA. 
However, for other targets  as $^{208}$Pb~\cite{Wooll04}, $^{27}$Al~\cite{Gomes04} and $^{89}$Y~\cite{pal14} the energy dependence do not present a clear trend.

\begin{figure}
 
\includegraphics[width=0.8\linewidth]{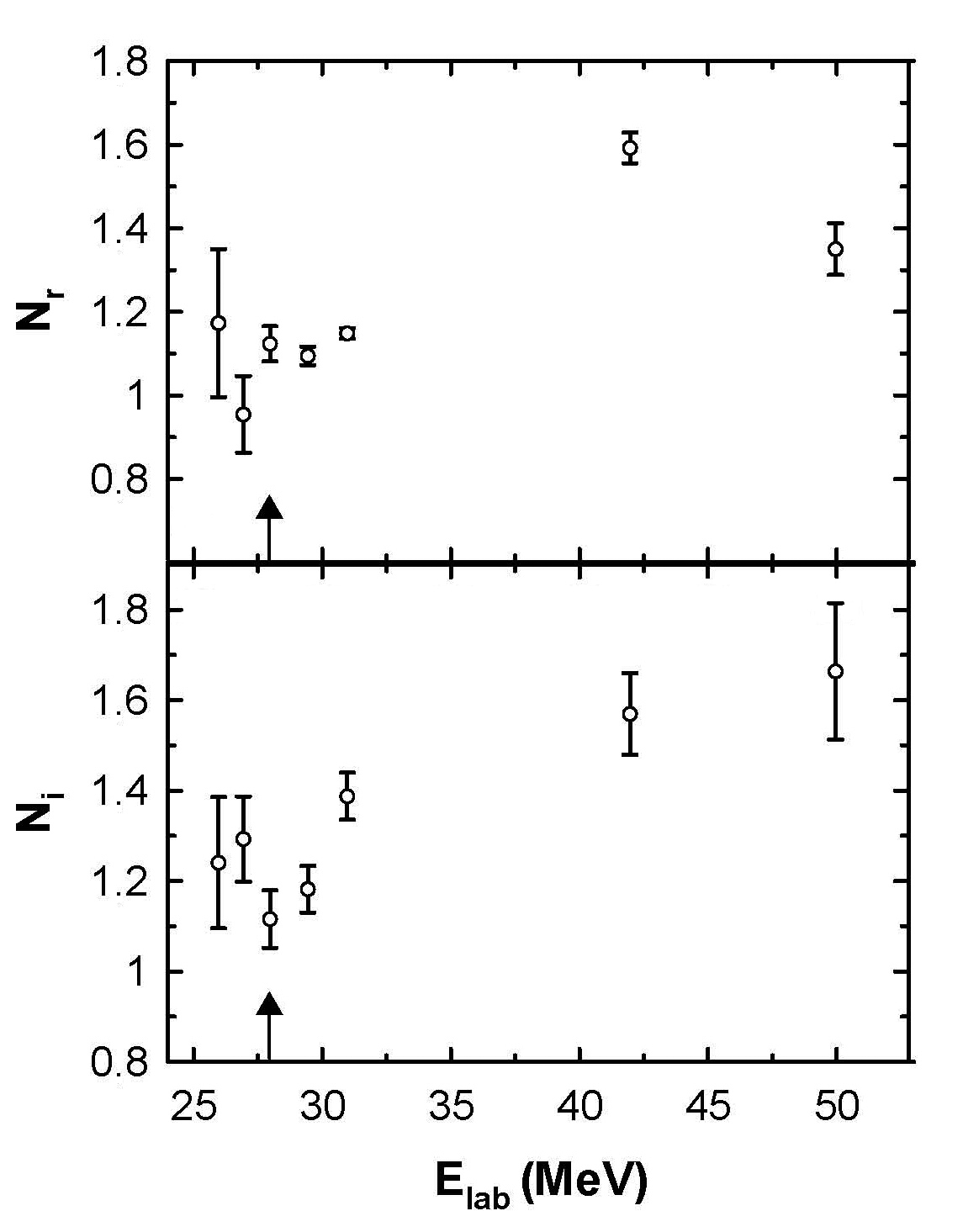}
\caption{Best real ($N_r$) and imaginary ($N_i$)  normalization factors for the
fitting of experimental elastic scattering angular distributions of the $^9$Be +
$^{120}$Sn system.  The uncertainty bars are calculated according to the procedure of Ref. \cite{abr15}. The vertical arrow shows the energy of the Coulomb barrier.}
\label{potenciales}
\end{figure}

  Within the range of bombarding energies studied in the present work (see Fig.  \ref{potenciales}), the $^9$Be$+^{120}$Sn system shows a slight decreasing trend of the
 imaginary factor $N_i$ at energies below the Coulomb barrier. However, this fall is not as pronounced as in the usual threshold anomaly presented by tightly bound projectiles. This can be interpreted as absorptive channels still being open at 
energies below the nominal Coulomb barrier ($V_{\rm C} \sim 28$~MeV). On the other hand, the expected break-up threshold anomaly (BTA), in which a rise of the imaginary term occurs before its  final drop, is also not found. Thus, the behaviour presented by the $^9$Be$+^{120}$Sn system seems to be closer to the anomaly presented by the weakly bound nucleus $^7$Li \cite{fig07, fig10}.

\begin{table}
    \centering
\begin{tabular}{ccccc}
  \toprule
$E_{lab}$ (MeV)       &     $N_r$   &    $N_i$   &   $N$ &   $\chi^2/\nu$  \\
  \colrule
\hline
26       &        1.17(18)  &       1.24(15)&        29             &   0.6    \\
27        &        0.95(9)        &    1.29(9)     &        41   &   0.5   \\
28        &        1.12(4)        &    1.12(6)     &        41   &   0.8   \\
29.5      &        1.09(2)        &   1.18(5)     &        41  &     1.4    \\
31        &        1.15(1)        &   1.39(5)     &        31   &    1.9  \\
42        &        1.59(4)        &    1.57(9)     &        28  &    1.4    \\
50        &        1.35(6)        &     1.66(15)    &        23  &    2.0   \\
\botrule
\end{tabular}

    \caption{Parameters of the optical model calculations using the S\~ao Paulo
potential: bombarding laboratory energy $E_{\rm lab}$, normalization factor for the real (imaginary) part $N_r$ ($N_i$), number of measurements $N$, and reduced $\chi^2$ value for the fitting ($\nu = N -2$ is the degree of freedom).  }
    \label{OP-parameters}
\end{table}

\begin{table}
    \centering
\begin{tabular}{cccc}
\toprule   
$E_{lab}$ (MeV)  & $\sigma^{OM}_R$(mb) & $\sigma_R^{CDCC}$(mb) &  $\sigma_{BU}^{CDCC}$(mb) \\
\colrule
26       &      130  & 191 &  38.4   \\
27        &     210  & 257 & 47.2 \\
28        &     307  & 351 & 57.2 \\
29.5      &     480  & 511 &  72.4 \\
31        &     682  & 669 & 86.1\\
42        &     1610 & 1478 &  126.2    \\
50        &     1980 & 1832 &  140.4 \\
\botrule
\end{tabular}

    \caption{ Total reaction cross sections obtained from the OM analyses ($\sigma^{OM}_R$) and from the four-body CDCC calculations ($\sigma_R^{CDCC}$) for all the bombarding energies. Predictions for the total breakup cross sections ($\sigma_{BU}^{CDCC}$) are also given.} 
    \label{trxsec}
\end{table}

  \subsection{Four-body CDCC calculations}

Loosely bound nuclei like $^9$Be are easily broken up into their constituents when
colliding with another nucleus. This effect can be properly treated within the CDCC
formalism~\cite{Yahiro86,Austern87}, including the coupling to the continuum part of
the spectrum. 
The scattering of $^9$Be on $^{120}$Sn can be described within the four-body CDCC
framework considering the projectile, $^9$Be, as a three-body system $(\alpha +
\alpha + n)$. The excitation of the target (as well as other possible channels like
core excitation or fusion) is included implicitly by the absorption due to the
optical potentials between the projectile fragments and the target.

To describe the states of the projectile, we use the pseudo-state method, which
consists in diagonalizing the Hamiltonian in a discrete basis of square-integrable
functions. Different bases have been used for three-body
systems~\cite{IJThompson04,Matsumoto06,MRoGa05,Descouvemont15}. Here we use, as in
Ref.~\cite{JCasal15}, the analytical transformed harmonic oscillator (THO)
basis~\cite{JCasal13}. We refer the reader to Ref.~\cite{JCasal14} for details about
the structure calculations for $^9$Be using the analytical THO basis. Then, 
the $^9$Be-$^{120}$Sn four-body wave functions are expanded in the internal states
of the three-body projectile, leading to a coupled-equations system that has to be
solved. For that, a multipole ($Q$) expansion is performed for each projectile
fragment-target interacting potential.
The procedure is explained in detail in Refs.~\cite{MRoGa08,JCasal15}.

The structure model for the three-body system $^9$Be includes two-body potentials
plus an effective three-body force. Since the three-body calculations are just an
approximation to the full many-body problem, the parameters of the three-body
potential are adjusted to reproduce the energy and matter radius of the ground state
($j=3/2^-$) and the energies of the known low-energy resonant states ($j=$1/2$^+$,
3/2$^+$, 1/2$^-$, and 5/2$^-$).
The $\alpha-n$ potential is taken from Ref.~\cite{IJthompson00} and the
$\alpha-\alpha$ potential is the
Ali-Bodmer interaction~\cite{AliBodmer}, modified to reproduce the experimental
phase shifts. 
These are shallow potentials in the sense that they include repulsive terms to
remove unphysical
two-body states.  The parameters of the analytical THO basis chosen are those used
also in Ref.~\cite{JCasal15}.  The maximum hypermomentum is set to $K_{\rm max}=10$
as in Ref.~\cite{JCasal15}, which has been checked to provide converged results
for reaction calculations at the range of energies considered. The convergence is
also  reached using a THO basis with $i_{\rm max}=8$ hyper\nobreakdash-radial excitations. The
calculated ground-state energy is $\varepsilon_B=-1.574$ MeV and rms matter radius
$r_{\rm mat}=2.466$ fm, to be compared to the experimental values  $\varepsilon^{\rm exp}_B=-1.5736$ MeV~\cite{Tilley04}  and $r^{\rm exp}_{\rm mat}=2.53$ fm~\cite{Liatard90}.

The interactions between each projectile-fragment and the target are represented by
an optical potential, including both Coulomb and nuclear contributions. The
$n-^{120}$Sn potential is represented by the Koning and Delaroche global
parametrization~\cite{KD} at the corresponding energy per nucleon. For the
$\alpha-^{120}$Sn interaction, we use the code by S. Kailas~\cite{Kailas}, which
provides optical model parameters for $\alpha$ particles using the results from
Ref.~\cite{Atzrott95}. Our model space includes $j^\pi=3/2^\pm, 1/2^\pm$ and
$5/2^\pm$ projectile states up to 8 MeV above the breakup threshold, which ensures
convergence of the elastic angular distributions for this reaction. The coupled
equations are solved up to $301/2$ partial waves, including continuum couplings to
all multipole orders, i.e., up to $Q=5$.

In Figs.~\ref{fig:all1} and \ref{fig:all2} we show the four-body CDCC calculations
at the different energies measured: 50, 42, 31, 29.5, 28, 27 and 26 MeV. 
Dashed lines are calculations including only the ground state of $^9$Be and solid
lines are the full CDCC calculations. In all cases, the agreement with the data is
improved when we include the coupling to the continuum part of the spectrum. These  couplings are important even at lower energies, where the inclusion of
breakup channels is essential to describe the experimental cross sections. 
 This result, together with Ref.~\cite{JCasal15}, in which  it is shown that the scattering on a light target at sub-barrier energies exhibits a much smaller coupling to breakup channels, confirms that Coulomb breakup is an important process at low energies. This is a consequence of the weakly-bound nature of $^9$Be. 

The agreement between the experimental data and the full four-body CDCC calculations
is overall quite good. However, at energies 29.5 and 28 MeV, the calculations
underestimate the data in the nuclear-Coulomb interference region (between 60$^\circ$ and
90$^\circ$, approximately). This effect has already been addressed for the reaction of
$^9$Be+$^{208}$Pb~\cite{Descouvemont15,JCasal15}, also at energies around the
Coulomb barrier for this system. 
 In principle, one expects that the scattering of a weakly bound
nucleus such as $^9$Be on a heavy target follows the same behavior reported, both experimentally and theoretically, for other
weakly bound nuclei such as $^6$He~\cite{san08,net10}, $^{11}$Li~\cite{cubero12}, $^{11}$Be~\cite{piet10,piet12,pes17}. All these
nuclei present a suppression of the rainbow at the interference
region when colliding with heavy targets, as the energy decreases down to or below the Coulomb barrier. This is due to the strong dipolar Coulomb coupling to the continuum states, although nuclear coupling can be also important~\cite{piet12,keeley10}.

Discrepancies in the nuclear-Coulomb interference region between the converged
calculations and the experiment,  in the present work and in Refs.~\cite{Descouvemont15,JCasal15},
could be attributed either to unaccounted systematic errors in  the experimental data  or to the theoretical models used.  Both model calculations~\cite{Descouvemont15,JCasal15} are consistent.
A better understanding of this issue requires, in addition to the elastic data, breakup angular distributions. The comparison between the elastic and breakup channels at the same angular region could clarify the situation and such an experiment is already being planned at the TANDAR Laboratory.

\begin{figure}[h]
 \includegraphics[width=0.9\linewidth]{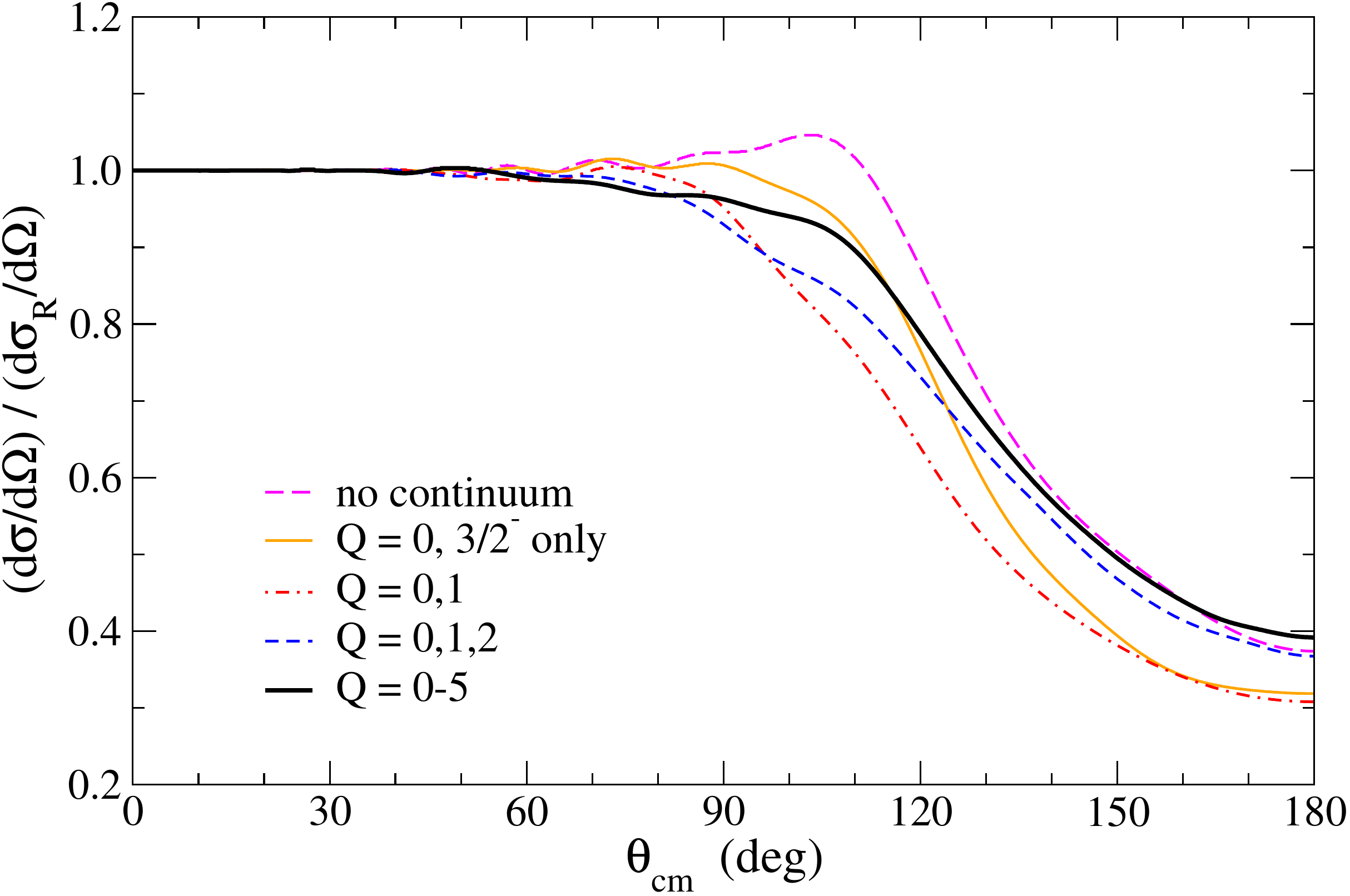} 
 \caption{(Color online) Angular distribution of the elastic cross section relative
to Rutherford for the reaction $^9$Be + $^{120}$Sn at $E_{\rm lab}=27$ MeV. The
effect of the different $j^\pi$ contribution and coupling multipolarities $Q$ is
shown.}
 \label{fig:Qdep}
\end{figure}

In order to study the effect of the $j^\pi$ contributions and coupling
multipolarities $Q$ on the results, in Fig.~\ref{fig:Qdep} we show different
calculations at $E_{\rm lab}=27$ MeV, i.e.~around the Coulomb barrier. The monopolar
($Q=0$) contribution allows to connect the 3/2$^-$ ground state to the 3/2$^-$
continuum. 
Then, dipolar and higher order terms introduce coupling between all $j^\pi$
configurations considered. 

We see in Fig.~\ref{fig:Qdep} that the main contributions to reduce the cross
section, the monopole and dipole terms, are of the same order. A similar result has
been reported previously for the scattering of $^9$Be on $^{208}$Pb~\cite{JCasal15}.
This result differs from the case of the scattering of halo nuclei on heavy targets,
e.g.~$^6$He or $^{11}$Li on $^{208}$Pb, where dipolar contributions produce the
largest deviation with respect to the calculation without continuum 
couplings~\cite{MRoGa08,cubero12}. The $^9$Be nucleus is a weakly-bound system but
presents a smaller $E1$ strength than typical halo nuclei~\cite{JCasal14}. This
reduces the effect of dipolar couplings. Higher order contributions also produce an
important effect which improves the description of the experimental data in the
whole angular region, but specially at backward angles.

 Last, given that we have no experimental breakup data, we compare in Table~\ref{trxsec}, the total reaction cross section for each energy as given for the four-body CDCC ($\sigma_R^{CDCC}$) calculations with the OM analyses ($\sigma^{OM}_R$). Both results are consistent considering that they come from very different approaches. Since the OM model is adjusted to the experimental data, this comparison supports the validity of the present CDCC calculations. These calculations also provide the total breakup cross sections ($\sigma_{BU}^{CDCC}$), which may serve as a prediction to guide future experiments on the breakup of $^9$Be on $^{120}$Sn.

\section{Target excitation: inelastic distributions}
\label{inel}

As stated in Sec. \ref{exp}, the experimental setup allowed to separate two
inelastic peaks on the spectra, from the elastic scattering peak, at the three
highest incident energies ($E_{\rm lab}=50$, 42 and 31 MeV). These peaks correspond to excitation energies of 1.19(5) MeV and 2.41(5) MeV above the g.s. The corresponding inelastic angular distributions are shown in Figs.~\ref{fig:inel1} and \ref{fig:inel2}, respectively.

\begin{figure}[h]
\includegraphics[width=0.9\linewidth]{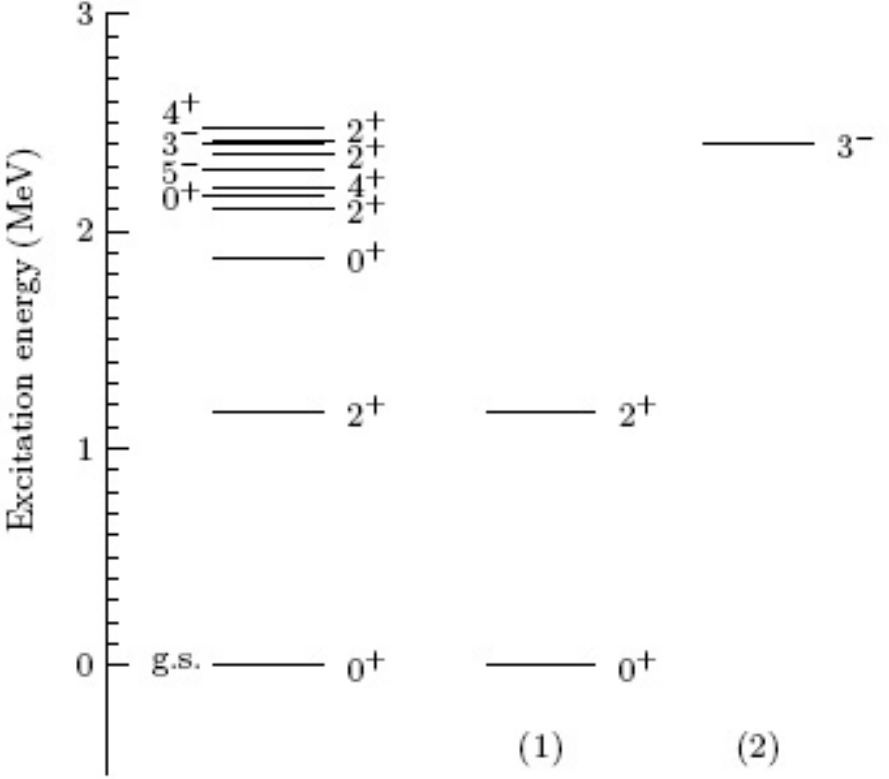} 
 \caption{ Low-energy states of the $^{120}$Sn nucleus. First column includes
the known experimental levels. Second and third columns are the first states of the
g.s. rotational band (1) and the negative parity band (2), respectively.}
 \label{fig:levels}
\end{figure}

Since $^9$Be has no bound excited states, these peaks are attributed to 
excitations of the  $^{120}$Sn target nucleus. Looking at the $^{120}$Sn known spectrum  
\cite{kit02}, the first excited state appears at an energy of 1.171265(15) MeV, over 
the ground state (g.s., $0^+$), with angular momentum $2^+$. Above the first excited 
state there are several states between 1.8 and 2.5 MeV (see Fig.~\ref{fig:levels}). 
According to the energy for which the second inelastic peak is observed, the states that can
contribute to this second peak are: $3^-$ at 2.40030(5) MeV, $2^+$ at 2.42090(3)
MeV, $2^+$ at 2.355383(24) MeV and $4^+$ at 2.465632(23) MeV.  Clearly, the experimental  energy resolution of the detectors (roughly about 200 keV) is not enough to distinguish individual contributions from these states to the second peak. 

To study the target excitations in the reaction of $^9$Be on $^{120}$Sn, we need a
 structure model for $^{120}$Sn. The nucleus of  $^{120}$Sn, and other even-even tin
isotopes~\cite{lee09}, do not exhibit neither typical rotational nor harmonic
vibrational structure.  The soft-rotator model \cite{porod96,chiba97} is usually
used to describe the collective level structure of this kind of nuclei. In
Ref.~\cite{lee09}, this model is used to sort the low-energy $^{120}$Sn states into,
approximately, rotational bands. Following the referred work, the g.s. ($0^+$) and
the first excited state ($2^+$ at 1.17 MeV over the g.s.) are members of the
so-called g.s. rotational band with $K\approx0$ as bandhead. The $3^-$ state at
2.40~MeV is the first level of the $K\approx0$ negative parity band.  The $2^+$ at
2.36~MeV is the second state of the gamma band with $K\approx0$ and was not included
in the subsequent reaction calculation. Finally, the states $2^+$ at 2.42~MeV and $4^+$ at
2.47~MeV are not even included in the structure calculation. According to this, the first inelastic
peak in the present work corresponds to the excitation to the first excited state
($2^+_1$ at 1.17 MeV, g.s. band) and the second peak to the first octupole
deformation state ($3_1^-$ at 2.40 MeV). This assumption is also supported by the
fact that $3_1^-$ is the only state, among the candidates, that has been detected
by Coulomb excitation \cite{kit02}.

Here, to analyze the experimental inelastic distributions, we perform simple
coupled-channels (CC) calculations with collective form factors~\cite{tamura65},
using matrix elements from a rigid rotor and taking the deformation parameters from
the literature. 
The quadrupole and octupole deformation parameters associated to the excitation of
the first 2$^+$ and 3$^-$ states, respectively, are taken as $\beta_2=0.1075$
\cite{raman01} and $\beta_3=0.1370$ \cite{kib02}.  From these values, the
calculated deformation lengths are 0.6363 and 0.8109 fm, respectively.
Apart from the deformation parameters, to perform the CC calculations is necessary
to introduce a
 bare potential between 
the projectile and the target, i.e., the interaction between
them in the absence of couplings to their internal degrees
of freedom. For each energy, we use here, as bare potential, the optical potential
obtained in Sec.~\ref{OMA} for the OM analysis of the elastic data at such energy.
The CC calculations were performed with the code {\sc FRESCO} \cite{thomp88}.

For the first excited state, the CC calculations are shown in Fig.~\ref{fig:inel1}
with a full line. The comparison between experimental data and CC calculations
is  very good, confirming the excitation to the 2$_1^+$ state in $^{120}$Sn. For the
second peak, the agreement is not so good, specially at the most backward angles
measured. In spite of the simplicity of the model calculation, these results
indicate that the second peak must be due, at least mostly, to the excitation of the
first octupole state $3^-_1$ at 2.40 MeV over g.s.

\section{Summary and conclusions}
\label{concl} 

We have measured the elastic scattering of the $^9$Be
nucleus on a $^{120}$Sn target at seven incident energies around and above the
Coulomb barrier ($E_{\rm lab}=50$, 42, 31, 29.5, 28, 27 and 26 MeV) at the TANDAR laboratory. In addition, the energy resolution of cooled silicon detectors allowed to separate
two inelastic peaks on the spectra, from the elastic scattering peak,  at the three
highest incident energies (50, 42 and 31 MeV) at excitation energies of 1.19(5)~MeV
and 2.41(5)~MeV.

The optical model analysis showed no significant drop of the absorption below the 
nominal Coulomb barrier, which can be interpreted as reaction channels being open 
for those energies. However, the appearance of a breakup threshold anomaly is not 
evident for this system.

The experimental elastic scattering distributions have been compared with four-body CDCC
calculations, describing the $^9$Be projectile as a three-body system
($\alpha+\alpha+n$). The overall agreement is quite good and the results show that 
the inclusion of the $^9$Be continuum is relevant for the scattering process even at energies around and below the Coulomb barrier.  This suggests that  breakup is important even at low energies.

Simple CC calculations with collective form factors, using matrix elements from a
rigid rotor, have been performed to confirm that the first inelastic peak  measured
corresponds to the excitation to the first excited state of the $^{120}$Sn nucleus,
$2^+_1$ at 1.17 MeV over the g.s. The calculations also suggest that the second
inelastic peak likely corresponds  to the octupole state $3_1^-$ at 2.40 MeV over
the g.s.

\begin{acknowledgments}

Authors are grateful to I.J. Thompson for his valuable support
concerning technical details with the code FRESCO. This work has been partially
supported by the Spanish Ministerio de Econom\'{\i}a y Competitividad and the
European Regional Development Fund (FEDER) under Projects  No. FIS2014-51941-P and
FIS2014-53448-c2-1-P, by Junta de Andaluc\'{\i}a under Group No. FQM-160 and Project No.
P11-FQM-7632 and by the European Union's Horizon 2020 research and innovation
program under grant agreement No. 654002. J. Casal acknowledges support from the
Ministerio de Educaci\'on, Cultura y Deporte, FPU Research Grant No. AP2010-3124. M.
Rodr\'{\i}guez-Gallardo acknowledges postdoctoral support from the Universidad de
Sevilla under the V Plan Propio de Investigaci\'on contract No. USE-11206-M.
R. Lichtent{\"a}hler acknowledges contract 2013/22100-7 from FAPESP (Brazil). 
Argentinean authors acknowledge grants PIP00786CO (CONICET) and PICT-2013-1363 (FONCyT).

\end{acknowledgments}

\bibliography{./bibfile}

\end{document}